\documentclass{elsart}

\usepackage{graphics}
\usepackage{amssymb}
 
\def\fq{{f^q}}

\def\be{{\begin{equation}}}
\def\ee{{\end{equation}}}

\begin{document}
\begin{frontmatter}

\title{Nonextensive statistical effects on the relativistic nuclear equation of state}

\author{A. Drago$^a$, A.~Lavagno$^{b}$, P.~Quarati$^{b}$}
\address{$^a$Dipartimento di Fisica, Universit\`a di Ferrara and INFN, 44100 Ferrara, 
Italy}
\address{$^b$Dipartimento di Fisica, Politecnico di Torino and INFN, 10129 Torino, 
Italy} 

\begin{abstract}
Following the basic prescriptions of the Tsallis' nonextensive thermodynamics, we 
study the relativistic nonextensive thermodynamics and the equation of state for a 
perfect gas at the equilibrium. The obtained results are used to study the relativistic 
nuclear equation of state in the hadronic and in the quark-gluon plasma phase. 
We show that small deviations from the standard extensive statistics imply remarkable 
effects into the shape of the equation of state. 
\end{abstract}

\end{frontmatter}

\section{Introduction}

It has been shown that 
the nonextensive generalization of the Boltzmann-Gibbs thermostatistics, proposed by 
Tsallis, can be very relevant in many physical applications where 
long-range interactions, long-range microscopic memories and/or fractal space-time 
constraints are present \cite{tsallis}. In particular, recently, many authors outline 
the possible connection to the nonextensive statistical framework with nuclear and 
high energy physical applications \cite{albe,wilk,rafe,bediaga,beck,ion}. 
The aim of this work is to generalize the basic concepts of the nonextensive 
statistical mechanics to the relativistic regime and to investigate, through the 
obtained relativistic thermodynamic relations, the relevance of nonextensive 
statistical effects on the hadronic and on the quark-gluon plasma (QGP) 
equation of state (EOS). As we will see, small deviations from the extensive 
thermostatistics produce a significant modification into the shape of the  
hadronic and QGP equation of state with important consequence on 
the deconfined phase transition and on several nuclear properties.

\section{Relativistic nonextensive thermodynamics}

In this section we present the basic macroscopic thermodynamics 
variables in the language
of the nonextensive relativistic kinetic theory. 
Let us start by introducing the particle four-flow in the phase space as \cite{lava}
\begin{equation}
N^\mu(x)=\frac{1}{Z_q}\int \frac{d^3p}{p^0} \, p^\mu \,f(x,p) \; ,
\label{nmu}
\end{equation}
and the energy-momentum flow as
\begin{equation}
T^{\mu\nu}(x)=\frac{1}{Z_q}\int \frac{d^3p}{p^0} \, p^\mu p^\nu \,
f^q(x,p) \; , 
\label{tmunu}
\end{equation}
where we have set $\hbar=c=1$, $x\equiv x^\mu=(t,{\bf x})$,
$p\equiv p^\mu=(p^0,{\bf p})$, $p^0=\sqrt{{\bf p}^2+m^2}$ is
the relativistic energy and $f(x,p)$ is the particle distribution function. 
The four-vector $N^\mu=(n,{\bf j})$ contains the probability density $n=n(x)$ 
(which is normalized to unity) and the probability flow ${\bf j}={\bf
j}(x)$. The energy-momentum tensor contains the normalized
$q$-mean expectation value of the energy density, as well as the energy flow,
the momentum and the momentum flow per particle. Its expression follows
directly from the definition of the mean $q$-expectation value in nonextensive 
statistics \cite{tsallis}; for this reason it is
given in terms of $f^q(x,p)$.

Furthermore, in the framework of the nonextensive thermostatistics, it
appears natural to generalize the nonextensive four-flow entropy 
$S_q^\mu(x)$ as follows
\begin{equation}
S_q^\mu(x)=-k_{_B} \,\int \frac{d^3p}{p^0} \,p^\mu f^q(x,p) [\ln_q
f(x,p)-1] \; . \label{entro4}
\end{equation}

It is possible to show \cite{lava} that such an entropy, 
together a generalized relativistic 
Boltzmann equation, satisfies the relativistic local $H$-theorem and
implies the following Tsallis-like equilibrium probability distribution
\begin{equation}
f_{eq}(p)= \frac{1}{Z_q}\left [1-(1-q) \frac{p^\mu U_\mu}{k_{_B}T}
\right]^{1/(1-q)} \; ,
\label{nrdistri}
\end{equation}
where $U_\mu$ is the hydrodynamic four-velocity \cite{groot} and
$f_{eq}$ depends only on the momentum in absence of an external
field. At this stage, $k_{_B}T$ is a free parameter and only in
the derivation of the equation of state it will be identified with
the physical temperature. 

We are able now to evaluate explicitly all other thermodynamic
variables and provide a complete macroscopic description of a
relativistic system at the equilibrium. Let us first calculate the
probability density defined as
\begin{equation}
n=N^\mu U_\mu=\frac{1}{Z_q}\int\frac{d^3p}{p^0} \, p^\mu U_\mu\,
f_{eq}(p)\;  .
\end{equation}
Since $n$ is a scalar, it can be evaluated in the rest frame where
$U^\mu=(1,0,0,0)$. Setting $\tau=p^0/k_{_B}T$ and $z=m/k_{_B}T$,
the above integral can be written as
\begin{equation}
n=\frac{4\pi}{Z_q} \, (k_{_B}T)^3 \int^\infty_z \!\!d\tau
(\tau^2-z^2)^{1/2} \, \tau \, e^{-\tau}_q \; .
\label{densrel}
\end{equation}
Let us introduce the $q$-modified Bessel function of the second 
kind as follows
\begin{equation}
K_n(q,z)=\frac{2^n n!}{(2n)!}\frac{1}{z^n}\int^\infty_z
\!\!d\tau (\tau^2-z^2)^{n-1/2}\,
\left(e^{-\tau}_q\right)^q \; , \label{qbessel}
\end{equation}
then, by means of a partial integration of Eq.(\ref{densrel}), 
the particle density can be cast into the compact form
\begin{equation}
n= \frac{4 \pi}{Z_q} \, m^2 \, k_{_B}T \, K_2(q,z) \;
.\label{densi}
\end{equation}

Considering the decomposition of the energy-momentum
tensor \cite{groot}: $T^{\mu\nu}=\epsilon\, U^\mu U^\nu-p\,
\Delta^{\mu\nu}$, where $\epsilon$ is the energy density, $p$ the
pressure and $\Delta^{\mu\nu}=g^{\mu\nu}-U^\mu U^\nu$, the
equilibrium pressure can be calculated as
\begin{equation}
p=-\frac{1}{3} T^{\mu\nu}\Delta_{\mu\nu}=-\frac{1}{3\,Z_q}
\int\frac{d^3p}{p^0} p^\mu p^\nu
\Delta_{\mu\nu}\fq_{\!\!\!\!\!eq}(p) \; ,
\label{pressrel}
\end{equation}
and can be expressed as
\begin{equation}
p=\frac{4\pi}{Z_q}\,m^2\,(k_{_B}T)^2\, K_2(q,z) \; . \label{press}
\end{equation}
Comparing Eq.(\ref{densi}) with Eq.(\ref{press}), we obtain 
\begin{equation}
p=n\,k_{_B}T  \; ,
\end{equation}
which is the equation of state of a perfect gas if we identify $T$
as the physical temperature of the system. 

We proceed now to calculate the energy density $\epsilon$ as 
\begin{equation}
\epsilon=T^{\mu\nu}U_\mu U_\nu=\frac{1}{Z_q}  \int\frac{d^3p}{p^0}
(p^\mu U_\mu)^2 \fq_{\!\!\!\!\!eq}(p) \; .
\label{energyrel}
\end{equation}
Inserting the previously defined variables $\tau$ and $z$ and
using the definition in Eq.(\ref{qbessel}), we obtain
\begin{equation}
\epsilon=\frac{4\pi}{Z_q}\, m^4 \left [
3\frac{K_2(q,z)}{z^2}+\frac{K_1(q,z)}{z}\right ] \; .
\end{equation}
Thus the energy per particle $e=\epsilon/n$ is
\begin{equation}
e=3 \,k_{_B}T +m \,\frac{K_1(q,z)}{K_2(q,z)} \; ,
\end{equation}
which has the same structure of the relativistic expression
obtained in the framework of the equilibrium Boltzmann-Gibbs
statistics \cite{groot}.

For a system of particles in degenerate regime the above classical distribution 
function (\ref{nrdistri}) has to be modified 
by including the fermion and boson quantum statistical prescriptions. 
For a dilute gas of particles and/or for small deviations from the standard 
extensive statistics ($q\approx 1$)  
the equilibrium distribution function, in the grand canonical ensemble, 
can be written as \cite{buyu}
\begin{equation}
n(k,\mu)=\frac{1}
{ [1+(q-1)(E(k)-\mu)/T ]^{1/(q-1)} \pm 1}  \, ,
\label{distrifd}
\end{equation}
where the sign $+$ stands for fermion and $-$ for boson particle.
All the previous results can be easily extended to the case of quantum statistical 
mechanics. 

\section{Nonextensive statistics in hadronic matter and QGP}

The motivation of the importance of non-standard statistical effects in nuclear and 
high energy physics lies in the fact the extreme conditions of 
density and temperature in high energy nuclear collisions give rise to 
memory effects and long--range color interactions. These conditions imply the presence 
of non--Markovian processes in the kinetic equation affecting the 
thermalization process toward equilibrium as well as the standard 
equilibrium distribution \cite{biro,ropke}. 
A rigorous determination of the conditions that produce a 
nonextensive behavior, due to memory effects and/or
long--range interactions,  should be based on microscopic calculations 
relative to the parton plasma originated during the high energy collisions. 
At this stage we limit ourselves to consider the problem from a qualitative 
point of view. However, it is noteworthy to notice that in proximity of the 
hadronic-QGP phase transition, non--perturbative QCD calculations become important. 
Only a small number of partons is present in the Debye sphere: the 
ordinary mean field approximation of the plasma is no longer correct 
and memory effects are not negligible. 
In addition, we observe that at high density the color magnetic 
field remains unscreened (in leading order) and long--range color magnetic 
interaction should be present at all temperatures.

From the above considerations it appears reasonable that in regime of high density and 
temperature both hadronic and quark-gluon EOSs must be affected by nonextensive 
statistical effects. In the next two subsections, we will study the two EOSs 
separately on the basis on the previously obtained relativistic thermodynamic relations.

\subsection{Nonextensive hadronic equation of state}

Concerning the hadronic phase we use a relativistic non-linear model based on 
the interacting many-particle system consisting of nucleons, isoscalar and isovector 
mesons ($\sigma$, $\omega$ and $\rho$)\cite{glen}. 
On the basis of the Eqs.(\ref{tmunu}), (\ref{pressrel}) and (\ref{energyrel}), 
the pressure and the energy density can be written as
\begin{eqnarray}
&P& =\sum_{i=n,p} \frac{2}{3}\int \frac{{\rm d}^3k}{(2\pi)^3}
\frac{k^2}{E_{i}^\star(k)}
[n^q(k,\mu_i^\star)+n^q(k,-\mu_i^\star)] -\frac{1}{2}m_\sigma^2\phi^2 \nonumber \\
&&\;\;\;\;\;\;\;\;\;- \frac{1}{3} a \sigma^3-\frac{1}{4}b\sigma^4+
\frac{1}{2}m_\omega^2\omega^2+\frac{1}{2}m_{\rho}^2 \rho^2,\\
&\epsilon&=
\sum_{i=n,p}{2}\int \frac{{\rm d}^3k}{(2\pi)^3}E_{i}^\star(k)
[n^q(k,\mu_i^\star)+n^q(k,-\mu_i^\star)]+\frac{1}{2}m_\sigma^2\sigma^2 \nonumber \\
&&\;\;\;\;\;\;\;\;\;+\frac{1}{3} a \sigma^3+\frac{1}{4}b\sigma^4
+\frac{1}{2}m_\omega^2\omega^2+\frac{1}{2}m_{\rho}^2 \rho^2,
\end{eqnarray}

where $n(k,\mu_i)$ and $n(k,-\mu_i)$ are the fermion particle and 
antiparticle distribution (\ref{distrifd}). 
The nucleon effective energy is defined as ${E_i}^\star=\sqrt{k^2+{{M_i}^\star}^2}$, 
where ${M_i}^\star=M_{N}-g_\sigma \sigma$.
The effective chemical potentials $\mu_i^\star$  are given in terms
of the vector meson mean fields 
${\mu_i}=\mu_i^\star  - g_\omega\omega\mp g_{\rho}\rho$ ($-$ proton, $+$ neutron),
where $\mu_i$ are the thermodynamical chemical potentials 
$\mu_i=\partial\epsilon/\partial\rho_i$. At zero temperature they
reduce to the Fermi energies $E_{Fi} \equiv \sqrt{k_{Fi}^2+{M_i^\star}^2}$ and the 
nonextensive statistical effects disappear. The isoscalar and isovector meson fields 
($\sigma$, $\omega$ and $\rho$) are obtained as a solution of the field equations 
in mean field approximation and the related couplings ($g_\sigma$, $g_\omega$ and 
$g_\rho$) are the free parameter of the model \cite{glen}.
Finally, The baryon densities $\rho_B$ are given by 
\begin{equation}
\rho_B=\gamma
\int\frac{{\rm d}^3k}{(2\pi)^3}[n(k,\mu_i^\star)-n(k,-\mu_i^\star)]\,,
\end{equation}
where $\gamma$ is the spin/isospin multiplicity. 

In Fig. 1, we report the obtained hadronic EOS: pressure as a function of the baryon 
number density (in units of the nuclear saturation density $\rho_0=0.16$ fm$^{-3}$) 
for different values of $q$. Because of previous phenomenological studies 
in heavy-ion collisions have brought to values of $q$ greater than unity \cite{albe}, 
we concentrate our analysis to $q>1$. The results are plotted at the temperature  
$T=100$ MeV and at fixed value of $Z/A=0.4$. The range of the considered baryon  
density and the chosen values of the parameters correspond to a physical 
situation that can realized in the recently proposed 
high energy heavy-ion collisions experiment(like in $Au-Au$ collisions) 
(see, for example, Ref.\cite{gsi}).

\begin{figure}[hbt]
\label{figbq}
\parbox{6cm}{ \scalebox{0.7}{
\includegraphics*[-30,550][500,800]{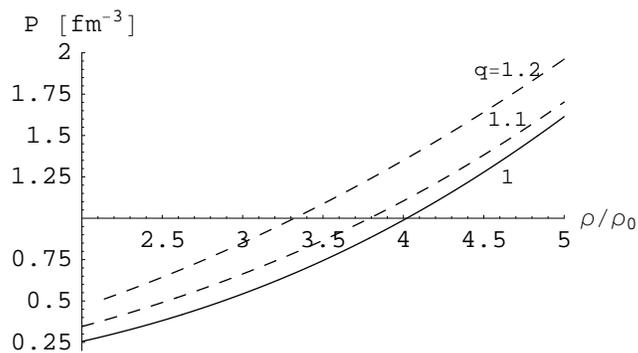}} }
\newline
\parbox{13cm}{
\caption{Hadronic equation of state: pressure versus baryon number density 
(in units of the nuclear saturation density $\rho_0$) for different values of $q$.}}
\end{figure}

\subsection{Nonextensive QGP equation of state}

In the simple model of free quarks in a bag, the pressure, energy 
density and baryon number density for a Fermi gas of quarks in the framework of 
nonextensive statistics (see Eqs.(\ref{nmu}), (\ref{tmunu}), (\ref{pressrel}) 
and (\ref{energyrel})) can be written, respectively, as 
\begin{eqnarray}
&P& =\sum_{f=u,d} \frac{1}{3}\frac{\gamma_f}{2\pi^2}
\int^\infty_0 k \frac{\partial\epsilon_f}{\partial k}
[n^q(k,\mu_f)+n^q(k,-\mu_f)]
k^2 dk -B, \\
&\epsilon& =\sum_{f=u,d} \frac{\gamma_f}{2\pi^2}
\int^\infty_0 \epsilon_f
[n^q(k,\mu_f)+n^q(k,-\mu_f)] k^2 dk+B, \\
&\rho& =\sum_{f=u,d} \frac{1}{3}\frac{\gamma_f}{2\pi^2}
\int^\infty_0 [n(k,\mu_f)-n(k,-\mu_f)] k^2 dk,
\end{eqnarray}
where $\epsilon_f=(k^2+m_f^2)^{1/2}$, $n(k,\mu_f)$ and $n(k,\mu_f)$ are 
the particle and 
antiparticle quark distributions. The quark degeneracy for each flavor is $\gamma_f=6$. 
Similar expressions for the pressure and the energy 
density can be written for the gluons as massless Bose gas with zero chemical 
potential and degeneracy factor $\gamma_g=16$. 
Because of the fermion-boson nonextensive distribution (\ref{distrifd}), 
the results are not analytical, also in the massless quark approximation. Therefore, 
a numerical evaluations of the integral must be performed. 
Note that a similar calculation,  
only for the quark-gluon phase, was also performed in Ref.\cite{miller} by studying 
the phase transition diagram.

In Fig. 2, we report the EOS for massless quark $u$, $d$ and gluons, 
for different values of $q$. As in Fig. 1, the results are plotted at the temperature  
$T=100$ MeV and at fixed value of $Z/A=0.4$; the bag parameter is $B^{1/4}$=170 MeV. 
In both the figures we can observe remarkable effects in the shape of the EOS for 
small deviations from the standard statistics.

\begin{figure}[hbt]
\label{figtr}
\parbox{6cm}{ \scalebox{0.7}{
\includegraphics*[-30,550][500,800]{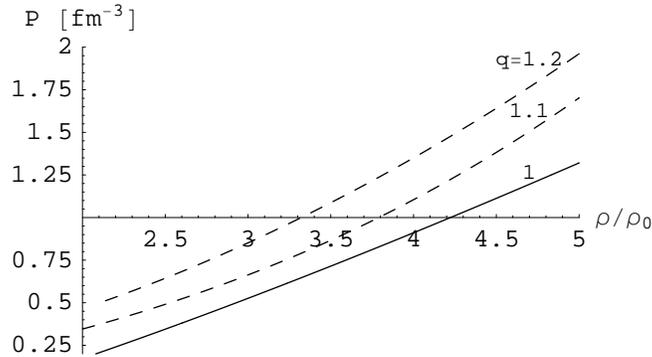}} }
\newline
\parbox{13cm}{
\caption{Same of Fig. 1 for the case of the quark-gluon equation of state 
with $B_1^{1/4}$=170 MeV.}}
\end{figure}

\section{Conclusion}

In this work we have studied the relativistic thermodynamic 
relations and derived the EOS of a gas of free particle at the equilibrium 
in the framework of the nonextensive Tsallis thermostatistics. The results are applied 
to obtain a consistent generalization of the EOS of strongly interacting 
hadronic matter and of deconfined QGP. 
The range of density and temperature chosen are physical values 
estimated in the recently proposed high energy 
heavy-ion collisions experiments at finite baryon chemical potential \cite{gsi}. 
We find that small deviations from the Boltzmann-Gibbs statistics implies a sensible 
modification of the two considered EOSs. A complete discussion on the nuclear physics 
consequence lies out the scope of this paper, however we want 
to outline that such a modification of the EOSs can strongly affect different physical 
properties, like, for example, the critical phase transition density, 
the symmetry energy, the nuclear compressibility, connected to experimental observables.

\end{document}